\documentstyle[12pt]{article}
\begin{document}
\baselineskip=15pt

\newcommand{\be}{\begin{equation}}
\newcommand{\ee}{\end{equation}}
\newcommand{\bq}{\begin{eqnarray}}
\newcommand{\eq}{\end{eqnarray}}
\newcommand{\x}{{\bf x}}
\newcommand{\p}{\varphi}
\newcommand{\Sc}{Schr\"odinger\,}
\newcommand{\del}{\nabla}
\newcommand{\A}{{\bf A}}
\newcommand{\nn}{\nonumber\\}

\newcommand{\bea}{\begin{eqnarray}}
\newcommand{\eea}{\end{eqnarray}}

\def\Journal#1#2#3#4{{#1} {\bf #2}, #3 (#4)}

\def\NPB{{\em Nucl. Phys.} B}
\def\PLB{{\em Phys. Lett.}  B}
\def\PRL{\em Phys. Rev. Lett.}
\def\PRD{{\em Phys. Rev.} D}

\begin{titlepage}
\rightline{DTP 96/33}
\vskip1in
\begin{center}
{\large A Large Distance Expansion for Quantum Field Theory} 
\end{center}
\vskip1in
\begin{center}
{\large
Paul Mansfield

Department of Mathematical Sciences

University of Durham

South Road

Durham, DH1 3LE, England}

{\it P.R.W.Mansfield@durham.ac.uk}
\end{center}
\vskip1in
\begin{abstract}
\noindent 
Using analyticity of the vacuum wave-functional under complex 
scalings, the
vacuum of a quantum field theory may be reconstructed from a derivative 
expansion valid for slowly varying fields. This enables the eigenvalue 
problem for the Hamiltonian to be reduced to algebraic equations. 
Applied to Yang-Mills theory this expansion leads to a confining force 
between quarks. 
\end{abstract}

\medskip
\centerline{\it Invited talk at the Second International 
Sakharov Conference on Physics}

\end{titlepage}

\section{\bf Introduction}
I will describe an approach to the eigenvalue problem for the Hamiltonian 
of a quantum field theory, $\hat H\,|E\rangle=E\,|E\rangle$,
in which states are constructed from their simple
large distance behaviour.\cite{Paul} This is in contrast to the usual 
approach to, say, Yang-Mills theory, which is built up from simple 
short-distance behaviour. For simplicity, I will concentrate on 
scalar field theory, although the results
also apply to Yang-Mills theory where the leading order
in the expansion which I will describe leads to an area law for the 
Wilson loop
\cite{Paul2} via a kind of dimensional reduction.\cite{halp}

In the \Sc representation 
the field operator, $\hat\varphi$, is diagonal and its conjugate
momentum represented by functional differentiation

\be
\langle \varphi |\,\hat\p (\x)=\langle \varphi |\,\hat\p (\x),\quad
\quad\langle \varphi |\,\hat\pi (\x)=-i{\delta\over\delta \p (\x)}\,
\langle \varphi |,
\ee
so
that the ground state is represented by the wave-functional
$\langle \varphi |E_0\rangle =\Psi [\p ]=\exp {W[\p]}.$
In general $W[\p]$ is {\it non-local}, but when 
$\p (\x)$ varies very slowly on length-scales that are large in comparison 
to the inverse of the mass of the lightest particle it has a derivative
expansion in terms of {\it local} functions, e.g. $W=\int d\x (a_1\p^2+
a_2\nabla\p\cdot\nabla\p+a_3\p^4..)$. This expansion is the basis of our method.
At first glance it would appear to be completely useless, because the 
internal structure of particles is characterised by much shorter scales,
although there is
one physically interesting phenomenon that takes place at arbitrarily 
large distances, the confinement of quarks. I will claim, however, that 
this 
large distance behaviour is relevant not just to confinement but to 
understanding physics on all length scales, because I will show how 
it may be used to reconstruct the wave-functional $\Psi[\p]$ for 
arbitrary $\p(\x)$.

\section{\bf The Local Expansion}

Consider $\langle \p|e^{-T\hat H}|\tilde\p\rangle$. According to Feynman 
this is given by an integral over fields $\phi (\x ,t)$ that live in a 
Euclidean space-time bounded by the surfaces $t=0$, and $t=-T$ on which 
$\phi$ is equal to $\p$ and $\tilde\p$ respectively. As $T\rightarrow 
\infty$ this matrix element is dominated by the contribution from the 
ground-state, so
\be
\langle \p|\,e^{-T\hat H}\,|\tilde\p\rangle=
\int {\cal D}\phi \, e^{-S_E}\sim \Psi [\p]\,e^{-TE_0}\,\Psi^*[\tilde\p]
=e^{W[\p]+W[\tilde\p]-E_0T},
\ee
where $S_E$ is the Euclidean action.
From this we can extract $\Psi[\p]$.
A different formulation makes the dependence on
$\p$ more explicit.\cite{sym} Define the bra $\langle D|$ so as to be 
annihilated by $\hat\p$, then 
the canonical commutation relations imply that $\langle \p |=
\langle D|\,\exp ({i\int d\x \,\hat\pi\p})$.
So now
\be
\langle \p|e^{-T\hat H}|\tilde\p\rangle=\langle D|\,e^{i\int d\x 
\,\hat\pi\p}\,e^{-T\hat H}\,e^{-i\int d\x \,\hat\pi\tilde\p}|D\rangle,
\ee
which can be written as the functional integral
\be
\int {\cal D}\phi \, e^{-S_E+\int d\x\,\dot\phi(\x,0)\,\p(\x)-\int d\x 
\,\dot\phi (\x,-T)\,\tilde\p (\x )}.
\label{eq:int}
\ee
The boundary condition on the integration variable, $\phi$, implied by 
$\langle D|$ is that it should vanish on the boundary surfaces 
$t=0$ and $t=-T$. (In replacing $\hat\pi$ by $\dot\phi$, the time 
derivative of $\phi$, we should also include  delta functions in time, 
coming from the time-ordering.) 
So $W[\p]$ is the sum of connected Euclidean Feynman diagrams in which 
$\p$ is a source for $\dot\phi$ on the boundary. The only major difference 
from the usual Feynman diagrams encountered in field theory is that the 
propagator vanishes when either of its arguments lies on the boundary. 
Using this, Symanzik discovered the remarkable result  that in $3+1$ 
dimensional $\phi^4$ theory $W[\p]$
is finite as the cut-off is removed. 
For a free scalar field with mass $m$ this gives $W=
-{1\over 2}\int d\x\,
\p\sqrt{-\nabla^2+m^2}\,\p$, so that if the Fourier transform of 
$\p$ vanishes
for momenta with magnitude greater than the mass, $W$ can be expanded 
in the convergent series 
$-\int d\x \left({m\over 2}\p^2+{1\over 4m}(\nabla \p)^2-
{1\over 16 m^3}
(\nabla^2\p)^2..\right)$.
The terms of this expansion are local 
in the sense that they involve the field and a finite number of 
its derivatives at the same spatial point. The same is true for an 
interacting 
theory in which the lightest particle has non-zero mass, 
because massive propagators are 
exponentially
damped at large distances so that configuration-space Feynman diagrams 
are negligible except when all their points are within a distance 
$\approx 1/m$ of each other.
Integrating these against slowly varying sources, $\p(\x)$, leads to 
local functions.

\section{\bf Reconstructing the Vacuum}
For $1+1$-dimensional scalar theory
define the scaled field $\p^s(x)=\p(x/\sqrt s)$ where $s$ is real and 
greater than zero. I will now show that $W[\p^s]$ extends to an analytic 
function of $s$
with singularities only on the negative real axis (at least within an 
expansion in powers of $\p$)
from which $W[\p]$ can be obtained using Cauchy's theorem. As 
$T\rightarrow\infty$ in (\ref{eq:int}), $\Psi$ becomes a functional 
integral on the Euclidean space-time $t\le 0$.  
By rotating the co-ordinates we can view this instead as a functional 
integral over the Euclidean space-time $x\ge 0$,
so
\be
e^{W[\p^s]}=
\int {\cal D}\phi \, e^{-S^{r}_E+\int dt\, \phi'(0,t)\p^s (t)},
\ee
where $\phi'=\partial\phi/\partial x$,
and $S^{r}_E$ is the action for the rotated space-time. This can be 
re-interpreted as
the time-ordered expectation value of $\exp \int dt\,( \p^s(t)\hat
\phi'(0,t)-\hat H^{r})$ in the ground-state, $|E^{r}\rangle$, of the 
rotated Hamiltonian, $\hat H^{r}$. The time integrals can be done if
this is
expanded in powers of $\p^s$, and the sources
Fourier analysed using $\tilde\p^s(k)=\sqrt s \tilde\p(k\sqrt s)$.
This yields
\bea
\Psi[\p^s]=
\sum_{n=0}^\infty \int dk_n..dk_1\,\tilde\p(k_n)&...&
\tilde\p(k_1)\,\delta({\textstyle\sum_1^n} k_i)\times\nonumber\\
{\sqrt s}^n\,\langle E^{r}_0|\hat\phi'(0){1\over  \sqrt s \hat H^{r}
+i({\textstyle\sum^{n-1}_1} k_i)}
\hat\phi'(0)&...&\hat\phi'(0){1\over \sqrt s \hat H^{r}+ik_1}
\hat\phi'(0)|E^{r}_0\rangle.
\eea
This can now be extended to the complex $s$-plane. Since the eigenvalues 
of $\hat H^{r}$ are real, the singularities occur for $s$ on the 
negative real axis. This must also hold for $W[\p^s]$, which is 
the connected part of $\Psi[\p^s]$, since any additional 
singularities could not cancel between connected and disconnected 
pieces. Now define
\be
I(\lambda)\equiv{1\over 2\pi i }\int_C {ds\over s-1}
e^{\lambda(s-1)}W[\p^s]
\ee
Where $C$ is a very large circle centred on the origin, beginning 
just below the negative real axis and ending just above. On $C$, 
$\p^s(x)=\p(x/\sqrt s)\approx \p(0)$ and so varies only very slowly 
with $x$, so here we can use our local expansion. Now collapse the 
contour to a small circle around $s=1$, which contributes $W[\p]$, 
and a contour, $C'$, surrounding the negative real axis. 
When $\Re (\lambda)>0$ the latter is exponentially suppressed (to 
check this note that for large $|s|$ we can use the local expansion, 
and elsewhere on $C'$ the integrand is bounded). 
Hence $W[\p]=\lim_{\,\Re (\lambda)\rightarrow\infty}I$, which is
expressed in terms of the local expansion only.
In practice 
we can truncate the series to a finite number of terms and work with a
large value of $\lambda$ to get a good approximation.

In the \Sc representation for $\p^4$ theory the term in the Hamiltonian
that needs to be regulated is $\int dx \,{\hat \pi}^2$. If we
introduce a momentum cut-off, $1/\epsilon$, and define $H_\epsilon$ as

\be
-{\textstyle{1\over2}}\int_{k^2<1/\epsilon} dk{\delta^2\over\delta
\tilde\p (k)
\delta\tilde\p (-k)}+\int dx\left({\textstyle{1\over 2}}(\p'^2+
M^2(\epsilon)\p^2)+{g\over 4!}\p^4-{\cal E}(\epsilon)\right),
\ee
where
$M^2$ and
${\cal E}$ are known functions that diverge as $\epsilon\downarrow 0$,
and $g$ and $E$ are finite, then the \Sc equation is $\lim_{\epsilon
\downarrow 0}(H_\epsilon-E)\Psi=0$. This cannot be applied directly 
to the local
expansion since the cut-off refers to short distances, whereas the 
local expansion is only valid at large distances. However, using the 
same technique as above, it may be shown that $(H_{s\epsilon}\Psi)
[\p^s]$ is analytic
in the $s$-plane with the negative real axis removed. The small-$s$ 
and 
large-$s$ behaviour are related by Cauchy's theorem, so
\be
\lim_{\lambda\rightarrow\infty}
{1\over 2\pi i }\int_C {ds\over s}e^{\lambda s}(\{H_{s\epsilon}-E\}
\Psi)[\p^s] =0
\ee
This leads to a separate equation for the coefficent of each 
independent local function of $\p$. A good approximation results from 
working to a finite order in $\lambda$, and taking $\lambda$ large, 
but finite.
Expanding in powers of $g$ reproduces standard perturbative results for 
short-distance phenomena, but these equations may also be solved without 
resorting to perturbation theory.  

\section*{Acknowledgments}
I would like to thank the Royal Society for a conference grant.
\vfill\eject

\end{document}